\documentclass[english,aps,manuscript,aps,manuscript]{revtex4}
\usepackage[T1]{fontenc}
\usepackage[latin9]{inputenc}
\usepackage{color}
\usepackage{float}
\usepackage{mathrsfs}
\usepackage{amsmath}
\usepackage{amssymb}
\usepackage{graphicx}

\makeatletter

\newcommand{\noun}[1]{\textsc{#1}}

\@ifundefined{textcolor}{}
{%
 \definecolor{BLACK}{gray}{0}
 \definecolor{WHITE}{gray}{1}
 \definecolor{RED}{rgb}{1,0,0}
 \definecolor{GREEN}{rgb}{0,1,0}
 \definecolor{BLUE}{rgb}{0,0,1}
 \definecolor{CYAN}{cmyk}{1,0,0,0}
 \definecolor{MAGENTA}{cmyk}{0,1,0,0}
 \definecolor{YELLOW}{cmyk}{0,0,1,0}
}

\makeatother

\usepackage{babel}
\begin{document}

\title{Dynamics of a solitonic vortex in an anisotropically trapped superfluid }

\author{J.M. Gomez Llorente and J. Plata}

\address{Departamento de F\'{\i}sica and IUdEA, Universidad de La Laguna,\\
 La Laguna E38200, Tenerife, Spain.}
\begin{abstract}
We analytically study the dynamics of a solitonic vortex (SV) in a
superfluid confined in a non-axisymmetric harmonic trap. The study
provides a framework for analyzing the role of the trap anisotropy
in the oscillation of SVs observed in recent experiments on atomic
Bose and Fermi superfluids. The emergence of common and statistics-dependent
features is traced in a unified approach to both types of fluid. Our
description, built in the hydrodynamic formalism, is based on a Lagragian
approach which incorporates the vortex location as dynamical parameters
of a variational ansatz. Previous operative Hamiltonian pictures are
recovered through a canonically traced procedure. \textcolor{black}{Our
results improve the understanding of the experimental findings. Some
of the observed features are shown to be specific to the tri-axial
anisotropy of the trap. In particular, we characterize the nontrivial
dependence of the oscillation frequency on the trapping transversal
to the vortical line. The study reveals also the crucial role played
by the nonlinear character of the dynamics in the observed oscillation:
for the considered experimental conditions, the frequency, and, in
turn, the effective inertial mass of the vortex, are found to significantly
depend on the amplitude of the generated motion. It is also uncovered
how the coupling with collective modes of the fluid induces a non-negligible
shift in the oscillation frequency. The appearance of }\textcolor{black}{\emph{fine-structure}}\textcolor{black}{{}
features in the SV }\textcolor{black}{\emph{trajectory}}\textcolor{black}{{}
is predicted.}
\end{abstract}
\maketitle

\section{Introduction}

The dynamics of solitonic excitations in trapped Bose-Einstein condensates
(BECs) have been the subject of intense research in the last decades
\cite{Phillips,Cornell-Snaking,Frantzeskakis,LewensteinInstability,Oberthaler,Sanpera,KonotopPitaevskii,Hau,Komineas1}.
A central objective of the theoretical work has been the characterization
of the effect of trapping on structures identified in uniform environments.
Significant advances have been made: different soliton-like solutions,
with topological forms depending on the confining characteristics,
have been found to the Gross-Pitaevskii (GP) equation. Moreover, the
occurrence, dependent on the trapping conditions, in particular, on
the effective dimensionality, of dynamical or energetic instability
has been traced. Instability-induced decay sequences connecting diverse
types of solutions have been described \cite{Komineas1,CarrBrand,MMateo1,AnglinSnaking,Clark,Fetter2}.
For specific confining properties, planar dark solitons (PDSs), vortex
rings (VRs), and SVs are present in a decay cascade initiated via
\emph{snake} instability \cite{CarrBrand,BrandReinhardt,MateoBrandChladni}.
Parallel advances have taken place in the experimental area: various
techniques have been implemented for the direct observation and the
controlled generation of the structures \cite{Phillips,Cornell-Snaking,Oberthaler,Sanpera,Hau,Steinhauer,PitaevskiiSVinBEC}.
The research, initially focused on bosonic fluids, has been extended
to superfluid Fermi gases \cite{Pitaevskii2,Zwierlein1,StringariRMPfermions,Zwierlein2,StringariPitaevskiiFermi,Zwierlein3},
where, the distinctive aspects of coherence and interactions imply
facing additional, fundamental and technical, problems. In particular,
the evaluation of the potential role of the fluid statistics in the
appearance of differential characteristics of the soliton-like excitations
is required. Here, we deal with some recently uncovered aspects of
this problem: experiments realized \textcolor{black}{by different
groups} have revealed nontrivial features of the dynamics of SVs which
seem to be common to fermionic and bosonic superfluids. In those experiments,
originally intended to the controlled production of PDSs in a fermionic
superfluid in the BEC-BCS crossover \cite{Zwierlein1,Zwierlein2,Zwierlein3}
and in a BEC \cite{PitaevskiiSVinBEC,DalfovoNaturePhys} \cite{Spielman1},
oscillating long-lived SVs were detected. In fact, the presence of
SVs was inferred from the characterization of the observed oscillatory
motion, specifically, from tracing unexpected large values of the
effective inertial masses. Subsequently, the conclusive identification
of the structures as SVs was achieved in the fermionic case through
the implementation of direct tomographic imaging. Tomographic techniques
also allowed observing instability cascades, where, in agreement with
the predictions, PDSs were found to decay into VRs, which, in turn,
evolved into SVs \cite{Zwierlein2}. \textcolor{black}{The SV character
of the structures was also corroborated in one of the bosonic implementations:
observed in free-expansion images, a twisted planar density depletion
around the vortex line and phase dislocations in the interference
pattern were identified as distinctive SV signatures \cite{PitaevskiiSVinBEC}.
Afterwards, a stroboscopic technique was used to monitor the real-time
dynamics \cite{DalfovoStroboscopic}. The analyses of the experiments
have incorporated numerical simulations based on the GP equation and
hydrodynamical approaches \cite{PitaevskiiSVinBEC,DalfovoEuropeanJP},\cite{Zwierlein1}.
With them, some of the experimental findings have been approximately
reproduced. Numerical results were also presented to support the applicability
of the technique of control implemented in \cite{Spielman1}. Despite
those achievements, additional work on the understanding of the experimental
results seems necessary. We will focus on three issues that require
further clarification. First, an accurate characterization of the
role of the trap anisotropy is needed: as emphasized in \cite{Spielman1},
the lack of models that incorporate the triaxial anisotropy present
in some of the practical setups implies that no precise reference
values of the oscillation frequencies are available (the associated
analyses were based on the approximate applicability of models set
up for axisymmetric trapping to the actual anisotropic confinement).
Second, removing the limitations of the linear approximation employed
in some of the descriptions is essential: the magnitude of the observed
amplitudes demands the nonlinear character of the dynamics to be explicitly
taken into account. It is worth stressing that, in the nonlinear regime,
since the oscillation frequency, and, in turn, the (effective) inertial
mass are amplitude dependent, their measured values cannot be identified
as intrinsic characteristics of the structures. Third, to achieve
a detailed characterization of the system dynamics, some restrictions
of the applied formalism must be overcome: although the use of effective
Hamiltonian approaches built from approximate expressions for the
free energy of the SV has served to understand salient aspects of
the dynamics, a more complete description, where second-order effects
can be included, is required.}  To deal with those issues, we generalize
 the approach presented in \cite{Fetter1} to describe the precession
of vortex lines in BECs. \textcolor{black}{In this seminal work, the
ansatz for the condensate wave function incorporates the phase of
a quantum vortex line and the ground-state Thomas-Fermi density. Additionally,
the vortex core is modeled by considering a zero-density region around
the vortex line with a width given by the healing length. Corrections
to such an ansatz will be estimated in the present work. Our results
will show the basic approach to be rather accurate.} We build a general
framework where the evolution of SVs in Bose or Fermi superfluids
can be analyzed. Focusing on the dynamics subsequent to the SV formation,
we will proceed by setting up a variational scheme where the vortex
position will be incorporated as dynamical parameters of the ansatz.
In this approach, diverse characteristics of the setups, like different
regimes of trap anisotropy or a broad range of oscillation amplitudes,
can be addressed. To account for second-order effects, the trial \emph{wave-function}
will be generalized along two lines. First, we will assess the role
of additional degrees of freedom of the vortex motion associated to
(potentially realizable) sets of initial conditions. Subsequently,
the coupling with collective modes of the fluid will be evaluated.
The resulting framework will enable us to improve the agreement with
the experimental results and predict the appearance of nontrivial
\emph{fine-structure} features.

The outline of the paper is as follows. In Sec. II, we present our
model system. The variational method used to characterize the dynamics
is introduced in Sec. III. As a proof of consistency, we connect with
previous operative approaches by presenting a completely traced application
of the Hamiltonian formalism. In Sec. IV, the general dynamical equations
are particularized to the cases of a BEC and of a superfluid Fermi
gas of atoms in the BEC-BCS crossover \cite{StringariPitaevskiiBook,ZwierleinKetterle,BCS-BEC}.
Additional information on the dynamics, extracted from the generalization
of the variational ansatz, is discussed in Sec. V. Some details of
the application of our approach to the considered experiments are
given in Sec. VI. Finally, the general conclusions are summarized
in Sec. VII.

\section{The model system}

We consider an atomic Bose or Fermi superfluid characterized by an
order parameter $\Psi(\mathbf{r},t)=\sqrt{\rho(\mathbf{r},t)}e^{iS(\mathbf{r},t)}$
{[}$\rho(\mathbf{r},t)$ and $S(\mathbf{r},t)$ are respectively the
density and phase of the fluid{]} which is assumed to obey the nonlinear
Schrödinger (NLS) equation

\begin{equation}
i\hbar\frac{\partial\Psi(\mathbf{r},t)}{\partial t}=\left[-\frac{\hbar^{2}}{2M}\boldsymbol{\nabla}^{2}+V_{ex}(\mathbf{r})+\mu[\rho(\mathbf{r},t)]\right]\Psi(\mathbf{r},t).\label{eq:TimeNonlinearSchr}
\end{equation}
The identification of the parameters in this equation depends on the
bosonic or fermionic character of the fluid. For a Bose fluid, $M$
denotes the mass $m_{A}$ of a condensate atom, and $\mu[\rho(\mathbf{r},t)]$
accounts for the interaction term, $g\rho$, $g$ being the coupling
strength. Eq. (\ref{eq:TimeNonlinearSchr}) corresponds then to the
GP equation \cite{StringariPitaevskiiBook}. On the other hand, for
a fermionic fluid \textcolor{black}{in the BEC-BCS crossover}, $M$
stands for the mass of a pair of atoms ($M=2m_{A}$) and the nonlinear
term incorporates the equation of state of the fluid which expresses
the chemical potential $\mu$ as a function of the density. Moreover,
assuming the applicability of the polytropic approximation \cite{Stringari1,Manini,WenWen},
the nonlinear term is written as $\mu[\rho(\mathbf{r},t)]=C\rho(\mathbf{r},t)^{\gamma}$,
where the polytropic index $\gamma$ is a characteristic of the (interaction-dependent)
fluid regime, and $C$ is a constant, which is usually expressed in
terms of reference values for the chemical potential and density.
In the BEC side of the crossover, the polytropic index takes the value
$\gamma=1$ {[}Eq. (\ref{eq:TimeNonlinearSchr}) can be again identified
with the GP equation{]}. Additionally, in the unitary regime \cite{Zwierlein1,ZwierleinKetterle,BCS-BEC},
the dynamics can be expected to be modeled by taking $\gamma=2/3$.
Since no formal differences exist between the description of the bosonic
case and that of the fermionic superfluid in the (molecular) BEC regime,
we will account for them in a unified way. 

Emulating the referred practical setups, a confining nonaxisymmetric
harmonic potential $V_{ex}(\mathbf{r})$ is considered, i.e., 

\begin{equation}
V_{ex}(x,y,z)=\frac{1}{2}\left(k_{x}x^{2}+k_{y}y^{2}+k_{z}z^{2}\right),\label{eq:ExternalPotential}
\end{equation}
 where $k_{i}$ ($i\equiv x,y,z$) denote the force constants of the
trap, the corresponding frequencies being $\omega_{i}=\sqrt{\frac{k_{i}}{m_{A}}}$.
As in the experimental arrangements \cite{Spielman1,Zwierlein1},
we consider cigar-shaped traps with transversal anisotropy, specifically,
it is assumed that $k_{y}>k_{x}\gg k_{z}$. The SVs were observed
to be oriented along the shortest confining direction ($OY$ axis),
which was argued to be a consequence of energetic instability. We
will tackle this issue farther on. Additionally, we will show that
the trap frequencies that directly affect the observed precession
frequencies are those transversal to SV orientation, i.e., $\omega_{x}$
and $\omega_{z}$.

To obtain stationary solutions to Eq. (\ref{eq:TimeNonlinearSchr}),
we make the change 

\begin{equation}
\Psi(\mathbf{r},t)=\sqrt{\rho(\mathbf{r})}e^{-\frac{i}{\hbar}\bar{\mu}t},\label{eq:StationaryOrderParameter}
\end{equation}
where $\bar{\mu}$ stands for the effective (bulk) chemical potential.
In the resulting (stationary) equation, the Thomas-Fermi (TF) approximation,
which corresponds to neglect the kinetic-energy term, implies making 

\begin{equation}
V_{ex}(\mathbf{r})+\mu[\rho(\mathbf{r})]=\bar{\mu}.\label{eq:TFapproximation}
\end{equation}
 Then, taking into account the polytropic approximation to $\mu[\rho(\mathbf{r})]$,
the stationary fluid density in the TF regime can be written as 

\begin{eqnarray}
\rho(\mathbf{r})=\left|\Psi(\mathbf{r})\right|^{2} & = & C^{-1/\gamma}\left[\bar{\mu}-V_{ex}(\mathbf{r})\right]^{1/\gamma}\nonumber \\
 & = & \rho_{0}\left[1-\frac{V_{ex}(\mathbf{r})}{\bar{\mu}}\right]^{1/\gamma},\label{eq:TFDensity}
\end{eqnarray}
 for $\bar{\mu}-V_{ex}(\mathbf{r})\geq0$, and, $\rho(\mathbf{r})=0,$
otherwise. {[}We have written $\rho(\mathbf{0})\equiv\rho_{0}$.{]}
The TF radii $R_{i}$, ($i\equiv x,y,z$), are given by 

\begin{equation}
R_{i}=\sqrt{\frac{2\bar{\mu}}{k_{i}}},\label{eq:TFradii}
\end{equation}
 $\bar{\mu}$ being obtained from normalization.

The requirements for the validity of the present approach must be
emphasized. First, the hydrodynamic description is expected  to be
valid at length scales much larger than the healing length. Second,
the above NLS equation {[}Eq. (\ref{eq:TimeNonlinearSchr}){]} is
applicable provided that a local density approximation to the equation
of state of the fluid is feasible. Those conditions are fulfilled
in the mentioned experiments.

\section{Description of the dynamics through a variational Lagragian approach}

\subsection{General dynamical equations}

The NLS equation given by Eq. (\ref{eq:TimeNonlinearSchr}) with $\mu[\rho(\mathbf{r},t)]=C\rho(\mathbf{r},t)^{\gamma}$
can be derived from the Lagragian density 

\begin{equation}
\mathcal{L}[\Psi]=i\frac{\hbar}{2}\left(\Psi^{*}\frac{\partial\Psi}{\partial t}-\Psi\frac{\partial\Psi^{*}}{\partial t}\right)-\frac{\hbar^{2}}{2M}\left|\boldsymbol{\nabla}\Psi\right|^{2}-V_{ex}\left|\Psi\right|^{2}-\frac{C}{\gamma+1}\left|\Psi\right|^{2(\gamma+1)}\label{eq:LagragianDensity}
\end{equation}
 Indeed, the NLS equation is the Euler-Lagrange equation that is obtained
by imposing the action 

\begin{equation}
\mathscr{S}[\Psi]=\int_{t_{1}}^{t_{2}}dt\int d\mathbf{r}\mathcal{L}[\Psi]\label{eq:Action}
\end{equation}
to be stationary against infinitesimal variations $\delta\Psi$ and
$\delta\Psi^{\ast}$ which fulfill $\delta\Psi(\mathbf{r},t_{1})=\delta\Psi(\mathbf{r},t_{2})=0,\,\forall\mathbf{r}$.
This Lagragian formalism provides us with an appropriate framework
for setting up a variational method \cite{Zoller1}. First, an ansatz
$\Psi(\mathbf{r},t;\mathbf{u})$ is proposed for the \emph{wave function}.
(A generic notation, $\mathbf{u}$, is used for the variational parameters.)
Then, by introducing the ansatz into Ec. (\ref{eq:LagragianDensity}),
and integrating, a Lagragian function is obtained, i.e., 

\begin{equation}
L(\mathbf{u},\mathbf{\dot{u}},t)=\int d\mathbf{r}\mathcal{L}\left[\Psi(\mathbf{r},t;\mathbf{u}),\Psi^{\ast}(\mathbf{r},t;\mathbf{u})\right].\label{eq:Integrated LagDensity}
\end{equation}
 Finally, the effective Lagrange's equations

\begin{equation}
\frac{d}{dt}\left(\frac{\partial L}{\partial\mathbf{\dot{u}}}\right)-\frac{\partial L}{\partial\mathbf{u}}=0,\label{eq:LagrangeEquations}
\end{equation}
 give the dynamics of the variational parameters, and, consequently,
the time evolution of $\Psi(\mathbf{r},t;u)$. 

In order to describe the dynamics of the considered solitonic vortex,
assumed to be aligned along the $y$ direction, we use the ansatz 

\begin{equation}
\Psi(\mathbf{r},t;x_{0},z_{0})=\left|\Psi(\mathbf{r})\right|e^{iS(x,z;t;x_{0},z_{0})},\label{eq:GenericAnsatz}
\end{equation}
 where the variational parameters, $x_{0}$, and $z_{0}$, correspond
to the time-dependent locatio\textcolor{black}{n $\mathbf{r_{0}}(t)=[x_{0}(t),z_{0}(t)]$
of the SV }which we intend to characterize. For the phase profile,
which must account for the circulating flow around the vortex, we
write 

\begin{eqnarray}
S(x,z,t;x_{0},z_{0}) & = & \arctan\left(\frac{x-x_{0}}{z-z_{0}}\right)-\frac{\bar{\mu}t}{\hbar}\nonumber \\
 & \equiv & S_{v}(x,z;x_{0},z_{0})-\frac{\bar{\mu}t}{\hbar}.\label{eq:FirstPhaseAnsatz}
\end{eqnarray}
 Additionally, for $\left|\Psi(\mathbf{r})\right|$, we take the background
Thomas-Fermi (TF) expression, given through Ec. (\ref{eq:TFDensity}):
as a first-order approximation, it is assumed here that the modification
of the background density due to the presence of the vortex has a
minor effect on the characterization of the parameter dynamics. Second-order
effects will be evaluated in Section V, where we will use a more elaborate
ansatz which incorporates changes in the density correlated with the
phase proposal and accounts for the potential interplay of the vortex
motion with collective modes of the fluid. \textcolor{black}{These
changes will also allow a more precise evaluation of the effects of
the vortex core.}

The dynamics of the parameters $x_{0}$ and $z_{0}$ are governed
by the Lagragian function obtained from Eq. (\ref{eq:Integrated LagDensity}),
i.e., 
\begin{equation}
L(x_{0},z_{0};\dot{x}_{0},\dot{z}_{0})=\int d\mathbf{r}[-\hbar\rho\frac{\partial(S_{v}-\frac{\bar{\mu}t}{\hbar})}{\partial t}-\frac{\hbar^{2}}{2M}(\left|\boldsymbol{\nabla}\rho^{1/2}\right|^{2}+\rho\left|\boldsymbol{\nabla}S\right|^{2})-V_{ex}\rho-\frac{C\rho^{\gamma+1}}{\gamma+1}]\label{eq:BIGlagragian}
\end{equation}
 where we have taken into account that 
\begin{equation}
i\frac{\hbar}{2}\left(\Psi^{*}\frac{\partial\Psi}{\partial t}-\Psi\frac{\partial\Psi^{*}}{\partial t}\right)=-\hbar\rho\frac{\partial S}{\partial t},\label{eq:LagSimplification1}
\end{equation}
 and, 

\begin{equation}
\left|\boldsymbol{\nabla}\Psi\right|^{2}=\left|\boldsymbol{\nabla}\rho^{1/2}+i\rho^{1/2}\boldsymbol{\nabla}S\right|^{2}.\label{eq:LagSimplification2}
\end{equation}
Given the form of the ansatz {[}see Eq. (\ref{eq:GenericAnsatz}){]}
which incorporates the variational parameters only through the phase
$S_{v}(x,z;x_{0},z_{0})$, it is apparent that the terms associated
with $\bar{\mu}$, $\left|\boldsymbol{\nabla}\rho^{1/2}\right|^{2}$,
$V_{ex}\rho$, and $\frac{C}{\gamma+1}\rho^{\gamma+1}$ in Eq. (\ref{eq:BIGlagragian})
are not relevant to the effective Lagrange's equations as  they do
not introduce dependence on the parameters. (Furthermore, since the
term $\left|\boldsymbol{\nabla}\rho^{1/2}\right|^{2}$ is neglected
in the considered TF approximation, it will yet be ignored in the
forthcoming generalization of the ansatz, in spite of the parameters
being incorporated then, not only through the phase, but also via
the density.) Hence, the Lagragian function can be effectively reduced
to
\begin{equation}
L(x_{0},z_{0};\dot{x}_{0},\dot{z}_{0})=\int d\mathbf{r}\rho\left[-\hbar\frac{\partial S_{v}}{\partial t}-\frac{\hbar^{2}}{2M}\left|\boldsymbol{\nabla}S_{v}\right|^{2}\right].\label{eq:SHORTlagragian}
\end{equation}
\textcolor{black}{The integration must exclude the region around the
solitonic-vortex core (a cylinder of radius equal to the healing length)
where the true condensate density can be accurately approximated to
zero. This approximation, which has been used, in all theoretical
previous works of confined vortex lines, regularizes the integral.}
Using the specific functional form proposed for the phase in Eq. (\ref{eq:FirstPhaseAnsatz}),
we rewrite Eq. (\ref{eq:SHORTlagragian}) as 

\begin{eqnarray}
L(x_{0},z_{0};\dot{x}_{0},\dot{z}_{0}) & = & \int d\mathbf{r}\rho(\mathbf{r})\frac{\hbar\left[\dot{x}_{0}(z-z_{0})-\dot{z}_{0}(x-x_{0})\right]-\frac{\hbar^{2}}{2M}}{(x-x_{0})^{2}+(z-z_{0})^{2}}\nonumber \\
 & \equiv & f_{x}(x_{0},z_{0})\dot{x}_{0}+f_{z}(x_{0},z_{0})\dot{z}_{0}+F(x_{0},z_{0}),\label{eq:CompactLagragian}
\end{eqnarray}
 where, for convenience for the later application of the Hamiltonian
formalism, we have introduced the functions

\begin{eqnarray}
f_{x}(x_{0},z_{0}) & = & \hbar\int d\mathbf{r}\rho(\mathbf{r})\frac{(z-z_{0})}{(x-x_{0})^{2}+(z-z_{0})^{2}},\label{eq:fxDEF}\\
f_{z}(x_{0},z_{0}) & = & -\hbar\int d\mathbf{r}\rho(\mathbf{r})\frac{(x-x_{0})}{(x-x_{0})^{2}+(z-z_{0})^{2}},\label{eq:fzDEF}\\
F(x_{0},z_{0}) & = & -\frac{\hbar^{2}}{2M}\int d\mathbf{r}\rho(\mathbf{r})\frac{1}{(x-x_{0})^{2}+(z-z_{0})^{2}},\label{eq:FfuncDEF}
\end{eqnarray}
 which have been evaluated using an approximate method of sequential
integration applicable in the regime of strong anisotropy corresponding
to the referred experiments, i.e., for $\omega_{x}\gg\omega_{z}$.
Here, we recall that, in the realization of \cite{Zwierlein1}, which
corresponded to a fermionic superfluid in the BEC-BCS crossover, the
displacement of the center of the (cigar-shaped) trap due to gravitational
effects led to a small difference between the values of the trap frequencies
in the directions perpendicular to the longest trap axis. Consequently,
\textcolor{black}{the system was not axially symmetric, the anisotropy
transversal to the vortical line being considerable}. Also, the trap
used in the experimental realization corresponding to an atomic BEC
\cite{Spielman1} was operated in a regime of strong anisotropy. We
have obtained for the above integrals 

\begin{eqnarray}
f_{x}(x_{0},z_{0}) & \simeq & 0,\label{eq:fxVALUE}\\
f_{z}(x_{0},z_{0}) & \simeq & -\pi\hbar\int_{-x_{0}}^{x_{0}}\rho_{2D}(x,z_{0})dx,\label{eq:fzVALUE}\\
F(x_{0},z_{0}) & \simeq & -\frac{\pi\hbar^{2}}{M}\rho_{2D}(x_{0},z_{0})\ln\left(\frac{R_{x}}{\xi}\right).\label{eq:FfunVALUE}
\end{eqnarray}
 where $R_{x}$ is the TF radius in the $x$ direction {[}see Eq.
(\ref{eq:TFradii}){]} and $\xi$ represents the size of the vortex
core, which, for, both, bosonic and fermionic fluids, can be approximated
as $\xi=\hbar/\sqrt{2M\bar{\mu}}$ within the TF picture. Note that
$\xi$ corresponds to the standard form of the healing length in the
bosonic case. Moreover, we have used the column density along the
vortex orientation, i.e., 

\begin{equation}
\rho_{2D}(x,z)=\int_{-y_{L}(x,z)}^{y_{L}(x,z)}\rho(x,y,z)dy.\label{eq:ColumnDensityDefinition}
\end{equation}
 where, 
\begin{equation}
y_{L}(x,z)=+\frac{1}{k_{y}^{1/2}}\left[2\bar{\mu}-\left(k_{x}x^{2}+k_{z}z^{2}\right)\right]^{1/2}\label{eq:ColumnDensityLimits}
\end{equation}
 is the limit value of the $y-$coordinate as a function of the other
two variables. We have employed (alternative) methods of integration
of general applicability to precisely define the range of validity
of the obtained Lagragian function. Namely, \textcolor{black}{although
exact values of the integrals present in Eqs. (\ref{eq:fxDEF}) and
(\ref{eq:fzDEF}) have not been explicitly obtained, they can be shown
to}\textcolor{green}{{} }lead to the same Lagragian function as the
expressions given by Eqs. (\ref{eq:fxVALUE}) and (\ref{eq:fzVALUE}).
Then, it follows that the approximation implemented to obtain $f_{x}(x_{0},z_{0})$
and $f_{z}(x_{0},z_{0})$ does not restrict the applicability of the
description. Additionally, in order to improve the accuracy of Eq.
(\ref{eq:FfunVALUE}), we have gone to the next precision order: we
have calculated the contribution of the zero-order terms (i.e, the
terms where the factor $R_{x}/\xi$ is not present). That contribution
will be incorporated through a numerical factor, the \emph{effective
zero-order parameter} $c_{\gamma}^{(0)}$, in the argument of the
logarithmic function in Eq. (\ref{eq:FfunVALUE}), namely, we will
write $\ln(c_{\gamma}^{(0)}R_{x}/\xi)$. In Sec. V, we will see that
the generalization of the description implies dealing with additional
zero-order terms, which will be accounted for by appropriately modifying
$c_{\gamma}^{(0)}$. The final value of that parameter, which entirely
incorporates the zero-order terms of the different extensions of the
model, will be given in Sec. VI, when the specific application of
the study to the experimental setups is discussed. 

Using the explicit forms of $f_{x}(x_{0},z_{0})$, $f_{z}(x_{0},z_{0})$,
and $F(x_{0},z_{0})$, the Lagragian function is written as 

\begin{eqnarray}
L(x_{0},z_{0};\dot{x}_{0},\dot{z}_{0}) & = & -\pi\hbar\dot{z}_{0}\int_{-x_{0}}^{x_{0}}\rho_{2D}(x,z_{0})dx-\pi\frac{\hbar^{2}}{M}\rho_{2D}(x_{0},z_{0})\ln\left(c_{\gamma}^{(0)}\frac{R_{x}}{\xi}\right),\label{eq:LagragianExplicitEx}
\end{eqnarray}
 and, from Lagrange's equations, one obtains that the evolution of
the vortex location is given by 
\begin{equation}
\dot{x}_{0}=\frac{\hbar}{2M}\frac{\frac{\partial\rho_{2D}(x_{0},z_{0})}{\partial z_{0}}}{\rho_{2D}(x_{0},z_{0})}\ln\left(c_{\gamma}^{(0)}\frac{R_{x}}{\xi}\right)\label{eq:DynEq1}
\end{equation}

\begin{equation}
\dot{z}_{0}=-\frac{\hbar}{2M}\frac{\frac{\partial\rho_{2D}(x_{0},z_{0})}{\partial x_{0}}}{\rho_{2D}(x_{0},z_{0})}\ln\left(c_{\gamma}^{(0)}\frac{R_{x}}{\xi}\right).\label{eq:DynEq2}
\end{equation}
 Here, it is apparent that much of the information on the dynamics
is incorporated into the column density. It is via $\rho_{2D}(x_{0},z_{0})$
that the anisotropy of the trap and the nonlinearity of the problem
enter the equations. Moreover, in the present approach, the differences
between the dynamics of the SV in bosonic and fermionic superfluids
emerge mainly from the different form of the column density in each
case. Later on, the specific functional form of $\rho_{2D}(x_{0},z_{0})$
will be introduced and the characteristic frequency of the oscillation
$\Omega_{p}$ will be obtained. We will see that the presence of the
quotient $(\partial\rho_{2D}/\partial z_{0})/\rho_{2D}$ {[}or $(\partial\rho_{2D}/\partial x_{0})/\rho_{2D}${]}
in the above equations implies that $\Omega_{p}$ does not explicitly
depend on the trap frequency along the SV direction $\omega_{y}$. 

Some considerations on dimensionality  are pertinent. The dynamical
system formed by the set of variational parameters has only one degree
of freedom: since the Lagrangian presents a linear dependence on the
generalized velocities, the dynamics are given by two first-order
equations. In consequence, only two initial conditions, e.g., the
vortex coordinates $x_{0}(t=0)$, $z_{0}(t=0)$, are required. Note
that the dimensionality constraints derive from the form chosen for
the variational ansatz. A reduction in the set of generalized coordinates
will be implemented in the next subsection.

\subsection{The Hamiltonian formalism}

An operative Hamiltonian picture set up from an approximate expression
for the free energy of the SV was presented in \cite{Zwierlein1}
and subsequently used in \cite{MMateo1}. (The same technique had
been applied to analyze the dynamics of a vortex ring in \cite{Pitaevskii1};
see also \cite{BulgacEarlyAttempt,MuellerEarlyAttempt} for alternative
approaches.) To establish the connection with those descriptions,
we give now a detailed account of the application of the Hamiltonian
formalism to our model system. 

Since building the Hamiltonian function from a redundant set of generalized
coordinates can lead to inconsistencies, we turn to implement a dimensionality
reduction, prior to the derivation of Hamilton equations. In order
to present a general procedure, explicit expressions for the functions
$f_{x}(x_{0},z_{0})$, $f_{z}(x_{0},z_{0})$, and $F(x_{0},z_{0})$
will not be used. We start by rewriting the Lagragian function that
governs the dynamics: to the expression of $L(x_{0},z_{0};\dot{x}_{0},\dot{z}_{0})$
given by Ec. (\ref{eq:CompactLagragian}), we add the total time derivative
of a function $G(x_{0},z_{0})$, which will be adjusted in order to
eliminate one of the generalized velocities. (Without loss of generality
we will remove $\dot{x}_{0}$.) Hence, the \emph{new} Lagragian function
is written as 

\begin{eqnarray}
\tilde{L}(x_{0},z_{0};\dot{x}_{0},\dot{z}_{0}) & = & f_{x}(x_{0},z_{0})\dot{x}_{0}+f_{z}(x_{0},z_{0})\dot{z}_{0}+F(x_{0},z_{0})+\frac{d}{dt}G(x_{0},z_{0})\nonumber \\
 & = & f_{x}(x_{0},z_{0})\dot{x}_{0}+f_{z}(x_{0},z_{0})\dot{z}_{0}+F(x_{0},z_{0})+\frac{\partial G}{\partial x_{0}}\dot{x}_{0}+\frac{\partial G}{\partial z_{0}}\dot{z}_{0},\label{eq:LagFunctRedDim}
\end{eqnarray}
 and, to eliminate $\dot{x}_{0}$, we impose

\begin{equation}
\frac{\partial G}{\partial x_{0}}=-f_{x}(x_{0},z_{0})\label{eq:GfunctionCondition}
\end{equation}
Then, $G(x_{0},z_{0})$ is obtained as

\begin{equation}
G(x_{0},z_{0})=-\int f_{x}(x_{0},z_{0})dx_{0},\label{eq:GfunctionDef}
\end{equation}
and the Lagragian function is converted into 

\begin{eqnarray*}
\tilde{L}(x_{0},z_{0};\dot{z}_{0}) & = & \left[f_{z}(x_{0},z_{0})+\frac{\partial G}{\partial z_{0}}\right]\dot{z}_{0}+F(x_{0},z_{0}).
\end{eqnarray*}
It follows that the canonical conjugate momentum of $z_{0}$ is given
by 

\begin{equation}
p_{z_{0}}=\frac{\partial\tilde{L}}{\partial\dot{z}_{0}}=f_{z}(x_{0},z_{0})+\frac{\partial G}{\partial z_{0}},\label{eq:canMomentum}
\end{equation}
and the Hamiltonian function is straightforwardly set up through a
Legendre transformation. Namely,

\begin{equation}
H(z_{0},p_{z_{0}})=p_{z_{0}}\dot{z}_{0}-\tilde{L}=-F\left[z_{0},x_{0}(z_{0},p_{z_{0}})\right]\label{eq:HamiltFreeEnergy}
\end{equation}
{[}We have written $x_{0}(z_{0},p_{z_{0}})$ from Ec. (\ref{eq:canMomentum}):
there are only two canonical conjugate variables, $z_{0}$ and $p_{z_{0}}$.{]}
The expression obtained for $H(z_{0},p_{z_{0}})$, particularized
to the axis-symmetric scenario, matches the effective Hamiltonian
function operatively built from the free energy in previous approaches
\cite{Zwierlein1} \cite{MMateo1}. Actually, in those descriptions,
the energy was evaluated from modeling the SV \emph{wave-function}
in a form analogous to the ansatz proposed in our variational method.
No \emph{ad hoc} introduction of the conjugate momentum is required
in our approach: a completely canonical procedure is followed.

Hamilton's equations are given by the expressions 

\begin{equation}
\dot{z}_{0}=\frac{\partial H}{\partial p_{z_{0}}}=-\frac{\partial F}{\partial x_{0}}\frac{\partial x_{0}}{\partial p_{z_{0}}},\label{eq:HamEq1}
\end{equation}

\begin{equation}
\dot{p}_{z_{0}}=-\frac{\partial H}{\partial z_{0}}=\left[\frac{\partial F}{\partial z_{0}}\right]_{x_{0}}+\frac{\partial F}{\partial x_{0}}\frac{\partial x_{0}}{\partial z_{0}},\label{eq:HamEq2}
\end{equation}
 which, after minor algebra, are shown to consistently reproduce the
dynamical equations obtained via the Lagragian formalism {[}Ecs. (\ref{eq:DynEq1})
and (\ref{eq:DynEq2}){]}. This alternative view of the evolution
of the parameters can be convenient for further studies where analogies
with other dynamical systems can be established via canonical transformations. 

At this point it is worth recalling that, in the referred experiments
\cite{Spielman1}, \cite{Zwierlein1}, the SVs were found to be oriented
along the shortest axis of the trap. Some insight into this finding
can be achieved by analyzing the expression of the Hamiltonian {[}Ec.
(\ref{eq:HamiltFreeEnergy}){]} for a generic orientation of the vortex.
It is shown that the lowest energy of the system corresponds indeed
to the vortical line oriented along the shortest radius. Consequently,
one can conjecture that there must be a damping mechanism which leads
to the occurrence of that minimum-energy orientation. (Dissipation
effects on related structures were studied in \cite{Shlyapnikov1,Shlyapnikov2,Spielman2,Lundh,LewensteinDissipation}.)
Closely connected with this aspect of the dynamics is the characterization
of the initial conditions for the analyzed process. In fact, this
is an open question: the possibility of preparing the initial state
is limited as the SVs seem to appear as the (uncontrolled) final product
of decay sequences which start with PDSs. Further restrictions on
the initial conditions are present if, as conjectured, the SVs experience
a damping process leading to the minimum-energy orientation. Actually,
from the potential effects of decay and damping, one can reasonably
expect the emergence of a constrained scenario for the effective preparation
of the SVs. The pertinence of a simple set of two initial conditions,
e.g., the coordinates of the vortical line, as required in the above
approach, seems to be corroborated by the general agreement of our
basic picture with the experimental results. On the other hand, a
potential realization where both, the initial positions and velocities
of the structures, could be independently fixed would require an approach
with a more elaborate ansatz where the dimensional reduction outlined
in the above paragraphs would not be feasible. We will deal with this
issue in Sec. V.

\section{The role of the fluid statistics in the vortex dynamics}

In our picture, the differential characteristics of the SV dynamics
in bosonic and fermionic superfluids are rooted in the corresponding
different values of the polytropic index of the applied NLS equation.
Furthermore, given the form of the ansatz used in the variational
method, $\gamma$ enters the description through the background density
$\rho(\mathbf{r})$, more specifically, via the column density $\rho_{2D}(x,z)$
and the effective factor $c_{\gamma}^{(0)}$ in Eqs. (\ref{eq:DynEq1})
and (\ref{eq:DynEq2}). To make explicit the differences between the
bosonic and fermionic cases, we evaluate the precise functional form
of that density: using the scaled variables 

\begin{equation}
X=\frac{x}{R_{x}},\quad Y=\frac{y}{R_{y}},\quad Z=\frac{z}{R_{z}},\label{eq:ScaledVariables}
\end{equation}
 the expression of the TF density in the bosonic case (i.e., for,
both, bosonic atoms and fermionic atoms in the BEC regime) is written
as 
\begin{equation}
\rho_{B}(X,Y,Z)=\rho_{0}\left(1-X^{2}-Y^{2}-Z^{2}\right).\label{eq:BosonicFinalDensity}
\end{equation}
 On the other hand, in the fermionic case at the unitarity regime,
the TF density reads 

\begin{equation}
\rho_{F}(X,Y,Z)=\rho_{0}\left(1-X^{2}-Y^{2}-Z^{2}\right)^{3/2}.\label{eq:FermFinalDens}
\end{equation}
 Both expressions are applicable in the range defined by $1\geq X^{2}+Y^{2}+Z^{2}$,
the density being zero outside that range. The respective column densities,
$\rho_{2D,B}(X,Z)$ and $\rho_{2D,F}(X,Z)$, are in turn given by 

\begin{equation}
\rho_{2D,B}(X,Z)=\frac{4}{3}R_{y}\rho_{0}\left(1-X^{2}-Z^{2}\right)^{3/2},\label{eq:BoseColDens}
\end{equation}
 and 

\begin{equation}
\rho_{2D,F}(X,Z)=\frac{3}{8}\pi R_{y}\rho_{0}\left(1-X^{2}-Z^{2}\right)^{2}.\label{eq:FermiColDens}
\end{equation}
Introducing those expressions into Ecs. (\ref{eq:DynEq1}) and (\ref{eq:DynEq2}),
we obtain the evolution of the vortex location, which can be expressed
in compact form, for both $\gamma=1$ and $\gamma=2/3$, as 

\begin{equation}
\dot{X}_{0}=-\frac{2\gamma^{-1}+1}{4}\frac{\hbar}{\bar{\mu}}\frac{\sqrt{k_{x}k_{z}}}{M}\ln\left(c_{\gamma}^{(0)}\frac{R_{x}}{\xi}\right)\frac{Z_{0}}{1-X_{0}^{2}-Z_{0}^{2}}\label{eq:VlocationEq1}
\end{equation}

\begin{equation}
\dot{Z}_{0}=\frac{2\gamma^{-1}+1}{4}\frac{\hbar}{\bar{\mu}}\frac{\sqrt{k_{x}k_{z}}}{M}\ln\left(c_{\gamma}^{(0)}\frac{R_{x}}{\xi}\right)\frac{X_{0}}{1-X_{0}^{2}-Z_{0}^{2}}\label{eq:VlocationEq2}
\end{equation}
 From these equations, it is readily shown that the amplitude of the
SV motion, given by 

\begin{equation}
A_{xz}=\sqrt{X_{0}^{2}+Z_{0}^{2}},
\end{equation}
 is a conserved magnitude satisfying $0\leq A_{xz}\leq1$. Indeed,
the vortex location describes the elliptical trajectory defined by
the equation 
\begin{equation}
\frac{x_{0}^{2}}{R_{x}^{2}}+\frac{z_{0}^{2}}{R_{z}^{2}}=A_{xz}^{2},\label{eq:Elipse-1}
\end{equation}
\textcolor{black}{which actually corresponds to the most conspicuous
experimental features. (Perturbative corrections to this description,
which will be presented in the next section, will allow us to predict
the emergence of fine-structure characteristics.)} Moreover, we can
combine Eqs. (\ref{eq:VlocationEq1}) and (\ref{eq:VlocationEq2}),
to obtain 

\begin{equation}
\ddot{X}_{0}+\Omega_{p}^{2}X_{0}=0,\label{eq:HarmonicX}
\end{equation}
 and 

\begin{equation}
\ddot{Z}_{0}+\Omega_{p}^{2}Z_{0}=0,\label{eq:HarmonicZ}
\end{equation}
 where the characteristic frequency of the vortex oscillation is 
\begin{equation}
\Omega_{p}=\frac{2\gamma^{-1}+1}{4}\frac{\hbar}{\bar{\mu}}\frac{\sqrt{k_{x}k_{z}}}{M(1-A_{xz}^{2})}\ln\left(c_{\gamma}^{(0)}\frac{R_{x}}{\xi}\right).\label{eq:OmegaPrec}
\end{equation}
In order to compare with the experimental results, it is convenient
to express $\Omega_{p}$ as a function of the trap frequencies $\omega_{i}=\sqrt{\frac{k_{i}}{m_{A}}}$,
($i\equiv x,y,z$). For a Bose gas of atoms, since $M=m_{A}$ and
$\gamma=1$, we find 

\begin{equation}
\Omega_{p,B}=\frac{3}{4}\frac{\hbar}{\bar{\mu}}\frac{\omega_{x}\omega_{z}}{(1-A_{xz}^{2})}\ln\left(c_{\gamma=1}^{(0)}\frac{R_{x}}{\xi}\right).\label{eq:OmPrecesionBose}
\end{equation}
In contrast, for a Fermi superfluid, taking into account that $M=2m_{A}$,
one obtains 

\begin{equation}
\Omega_{p,F}=\frac{2\gamma^{-1}+1}{8}\frac{\hbar}{\bar{\mu}}\frac{\omega_{x}\omega_{z}}{(1-A_{xz}^{2})}\ln\left(c_{\gamma}^{(0)}\frac{R_{x}}{\xi}\right).\label{eq:OmPrecFermi}
\end{equation}

Some preliminary clues to clarify the experimental findings can be
extracted from the above picture: 

i) The lack of models strictly applicable to a triaxial anisotropic
confinement was a handicap in the early interpretation of the experimental
results. In fact, former analyses were based on assuming the approximate
applicability of theoretical results known for an axisymmetric trapping.
From the initial interpretation of the findings as reflecting the
effect of the transversal trapping on the reduced mono-dimensional
motion of the observed structure, the oscillation frequency was conjectured
to depend on the trap frequencies transversal to the longest radius
of the (cigar-shaped) trap. In order to reproduce the observed features,
an effective mean value of the two transversal frequencies, tentatively
defined in different forms, was used in the expression known for the
axisymmetric setting \cite{Spielman1}. Moreover, an effective transversal
radius $R_{t}$ was incorporated in the logarithmic factor present
in the functional form of the frequency, i.e., $\ln\left(R_{t}/\xi\right)$.
Those limitations of the analysis are removed by the present study.
Our results conclusively show that it is the trap anisotropy transversal
to the vortex line that affects the precession frequency: as shown
in Eqs. (\ref{eq:OmPrecesionBose}) and (\ref{eq:OmPrecFermi}), $\Omega_{p}$
depends on both $\omega_{x}$ and $\omega_{z}$. Moreover, as we have
previously stated, $\Omega_{p}$ does not explicitly depend on the
trap frequency corresponding to the direction of the vortical line:
$\omega_{y}$ enters Eqs. (\ref{eq:OmPrecesionBose}) and (\ref{eq:OmPrecFermi})
only through the bulk chemical potential. Our study lifts also the
ambiguity relative to the argument of the logarithmic function: it
is the radius $R_{x}$ corresponding to the shortest transversal direction
to the vortex line that enters that argument.

ii) Because of the nonlinear character of the dynamics, the oscillation
frequency depends on the amplitude $A_{xz}$. In fact, for the amplitudes
detected in the experiments, a linear approximation, i.e., taking
$1-A_{xz}^{2}=1-X_{0}^{2}-Z_{0}^{2}\simeq1$, is not feasible, and
the specific value of the factor $1-A_{xz}^{2}$ must be taken into
account to reproduce the measured frequencies. 

A comment on the use of an effective inertial mass in the present
context is in order. The introduction of that concept in the study
of planar solitons \cite{Pitaevskii3} allowed deriving a compact
expression for the period of the soliton in terms of the period of
the (elongated) trap. In the considered regime of small-amplitude
oscillations, the inertial mass  is an intrinsic characteristic of
the (trapped) solitonic structure. However, in its application to
the (two-dimensional) dynamics of the SV made in \cite{Zwierlein1},
the inertial mass becomes dependent on the vortex position via the
column density. The present analysis shows that there is an additional
dependence on the SV position associated to the nonlinearity of the
dynamics. 

iii) No qualitative differences in the SV dynamics for the bosonic
and fermionic cases are predicted with the used variational framework.
The only differential effect is a numerical factor determined by the
value of the polytropic index $\gamma$ in the characteristic frequency
of oscillation $\varOmega_{p}$ \textcolor{black}{and the logarithmic
factor $c$}. The common global properties simply derive from the
assumed superfluid character of both Bose and Fermi gases.

\section{Generalization of the approach }

In this Section, the above description will be generalized by increasing
the flexibility of the variational ansatz. Two lines will be followed.
First, we will use a trial \emph{wave-function} where the vortex-location
parameters will be incorporated, not only through the phase, as in
the previous approach, but also via the functional form of the density.
This will be shown to imply dealing with additional degrees of freedom
in the characterization of the vortex dynamics. In the second line,
an ansatz which can account for the role of the condensate degrees
of freedom will be employed. Although both extensions can be studied
simultaneously, we will deal with them consecutively in order to concentrate
on their differential implications. For simplicity, only the analysis
of SV dynamics in the bosonic superfluid will be presented. For the
fermionic case, which can be straightforwardly studied following the
same procedure, only the final results will be given.

\subsection{Effects associated to vortex-induced variations in the fluid density}

Here, we use the connection between phase and density given by the
Euler-like equation of the hydrodynamic formalism \cite{StringariPitaevskiiBook}
to derive the functional form of the density from the form proposed
for the phase. Specifically, in the trial \emph{wave-function}, written
now as $\Psi=\tilde{\rho}^{1/2}e^{iS}$, our proposal for the phase
is 

\begin{equation}
S(\mathbf{r},t;x_{0},z_{0})=S_{v}(x,z;x_{0},z_{0})-\frac{1}{\hbar}\bar{\mu}t+\delta,\label{eq:PhaseAnstaz1Gen}
\end{equation}
 which still incorporates the \emph{vorticity-conveying} function
$S_{v}(x,z;x_{0},z_{0})$, given by Eq. (\ref{eq:FirstPhaseAnsatz}),
and the term associated to the bulk chemical potential $\bar{\mu}$.
So there are no differences with the previous proposal except for
the presence of the additional parameter $\delta(t)$, needed now
to guarantee the normalization of the trial \emph{wave-function }as
modifications in the density are introduced. Indeed, the form of the
density, $\tilde{\rho}$, is not longer assumed to be that of the
background $\rho(\mathbf{r})=g^{-1}\left[\bar{\mu}-V_{ex}(\mathbf{r})\right]$.
\textcolor{black}{Now, $\tilde{\rho}$ is left as a free field in
the Lagrangian density, it being subsequently fixed by its own Euler-Lagrange
equation of motion. It amounts to use the precise connection between
density and phase} given by the Euler-like (hydrodynamic) equation,
i.e., 

\begin{equation}
\frac{\hbar^{2}}{2M}\left|\boldsymbol{\nabla}S\right|^{2}+V_{ex}+\hbar\frac{\partial(S_{v}-\frac{\bar{\mu}t}{\hbar}+\delta)}{\partial t}+g\tilde{\rho}=0.\label{eq:Eulerlike}
\end{equation}
 Accordingly, the density is obtained in terms of the phase as

\begin{eqnarray}
\tilde{\rho} & = & -\frac{1}{g}\left[\frac{\hbar^{2}}{2M}\left|\boldsymbol{\nabla}S\right|^{2}+V_{ex}+\hbar\frac{\partial S_{v}}{\partial t}-\bar{\mu}+\hbar\dot{\delta}\right]\nonumber \\
 & = & \rho-\frac{1}{g}\left[\frac{\hbar^{2}}{2M}\left|\boldsymbol{\nabla}S\right|^{2}+\hbar\frac{\partial S_{v}}{\partial t}+\hbar\dot{\delta}\right].\label{eq:DensityEuler1}
\end{eqnarray}
 \textcolor{black}{This expression is incorporated now into our variational
scheme. To derive the Lagragian function, we first introduce into}
Eq. (\ref{eq:BIGlagragian}) the form of the nonlinear term corresponding
to the considered bosonic case. \textcolor{black}{It is worth emphasizing
that a more complete characterization of the vortex core is given
in this approach: the kinetic term, $\frac{\hbar^{2}}{2M}\left|\boldsymbol{\nabla}S\right|^{2}$,
explicitly accounts for the density reduction in the core. The size
of the zero-density region is found to correspond to the healing length,
as was assumed in the previous simpler model. Actually, the predictions
of the present approach confirms the applicability of the basic model
as a first-order approximation.} Then, using the link between phase
and density given by Eq. (\ref{eq:Eulerlike}), the Lagragian function
is written in the compact form 

\begin{equation}
L(x_{0},z_{0};\dot{x}_{0},\dot{z}_{0})=\frac{g}{2}\int d\mathbf{r}\tilde{\rho}^{2}.\label{eq:LagFirstGen}
\end{equation}
 Finally, by inserting in this equation the density as given by Eq.
(\ref{eq:DensityEuler1}), and retaining only the dominant terms,
we arrive at 

\begin{eqnarray}
L(x_{0},z_{0};\dot{x}_{0},\dot{z}_{0}) & = & \frac{g}{2}\int d\mathbf{r}\rho^{2}+\nonumber \\
 &  & \int d\mathbf{r}\rho\left[-\hbar\frac{\partial S_{v}}{\partial t}-\frac{\hbar^{2}}{2M}\left|\boldsymbol{\nabla}S_{v}\right|^{2}\right]+\nonumber \\
 &  & \frac{1}{2g}\int d\mathbf{r}\left(\hbar\frac{\partial S_{v}}{\partial t}\right)^{2}+\nonumber \\
 &  & \frac{1}{2g}\int d\mathbf{r}\left(\frac{\hbar^{2}}{2M}\left|\boldsymbol{\nabla}S\right|^{2}\right)^{2}\label{eq:FirstLagExt}
\end{eqnarray}
 The magnitude of the terms left out in the above equation can be
shown to be much smaller than that of the ones kept. We have also
omitted the part that accounts for the dynamics of $\delta(t)$, which
is uncoupled from the rest of parameters and it is trivially solved
in the regime which will be eventually considered. The term $\frac{g}{2}\int d\mathbf{r}\rho^{2}$
in the above equation can be ignored as it does not contain the variational
parameters. The rest of the integrals are evaluated to logarithmic
accuracy, including zero-order terms, to give

\begin{eqnarray}
L(x_{0},z_{0};\dot{x}_{0},\dot{z}_{0}) & = & -\pi\hbar\dot{z}_{0}\int_{-x_{0}}^{x_{0}}\rho_{2D}(x,z_{0})dx-\pi\frac{\hbar^{2}}{M}\rho_{2D}(x_{0},z_{0})\ln\left(c_{\gamma=1}^{(1)}\frac{R_{x}}{\xi}\right)+\nonumber \\
 &  & \pi\frac{\hbar^{2}}{g}\sqrt{\frac{2\bar{\mu}}{M\omega_{y}^{2}}}\ln\left(c_{\gamma=1}^{(2)}\frac{R_{x}}{\xi}\right)(\dot{x}_{0}^{2}+\dot{z}_{0}^{2}).\label{eq:LagFgenExp}
\end{eqnarray}
The first line contains the Lagragian function used in the former
approximation. Specific to the modification of the ansatz is the change
in the effective zero-order parameter, i.e., $c_{\gamma=1}^{(0)}\rightarrow c_{\gamma=1}^{(1)}$,
which is made to account for the contribution of the last integral
in Eq. (\ref{eq:FirstLagExt}), entirely given by zero-order terms.
Also emergent is the quadratic function of the generalized velocities
present in the second line. \textcolor{black}{(The parameter $c_{\gamma=1}^{(2)}$
is required there.)} The magnitude of the changes, which can be approximately
evaluated using the dynamical equations of the previous order of approximation,
i.e., Eqs. (\ref{eq:VlocationEq1}) and (\ref{eq:VlocationEq2}),
is shown to be much smaller than that of the former Lagragian. As
a consequence, the effects incorporated by the implemented modification
of the trial \emph{wave-function} can be estimated to correspond to
a perturbation of the formerly characterized scenario. \textcolor{black}{Particularly
relevant to the consistency of the description is the explicit inclusion
of the vortex core in the ansatz used in the modified scenario. The
minor effect of this correction justifies the approach followed in
previous theoretical works. Namely, the use of an ansatz where the
vortex core is modeled by an exclusion region in the background Thomas-Fermi
density with size determined by the healing length correctly accounts
for the most conspicuous experimental features. }

Some more specific conclusions follow:

i) Since the Lagragian presents a quadratic dependence on the generalized
velocities, the dynamics are described now in terms of two second-order
equations. Hence, no dimensional reduction can be implemented: both,
the initial positions and velocities, are needed to integrate the
equations. This approach can then be relevant to potential experimental
arrangements where the independent variation of that set of initial
conditions could be feasible. 

ii) As can be shown from the dynamical equations, the amplitude $A_{xz}=\sqrt{X_{0}^{2}+Z_{0}^{2}}$
is not longer a conserved magnitude. Indeed, the amplitude is found
to oscillate: Fig. 1 illustrates how the elliptic trajectories of
the vortex location found in the previous approach are now modulated
by a term oscillating with a frequency larger than the precession
frequency. 

iii) Useful insight into the mechanisms responsible for the dynamics
is given by a linear approximation. One of the normal modes can be
basically traced to a linearized version the model system of the previous
order of approximation. Its frequency, i.e., the (linear) counterpart
of the precession frequency formerly obtained, is approximately given
by 

\begin{equation}
\Omega_{p,B}=\frac{3}{4}\frac{\hbar}{\bar{\mu}}\frac{\sqrt{k_{x}k_{z}}}{M}\ln\left(c_{\gamma=1}^{(1)}\frac{R_{x}}{\xi}\right),\label{eq:OmegaPrec-1}
\end{equation}
 and 

\begin{equation}
\Omega_{p,F}=\frac{\hbar}{\bar{\mu}}\frac{\sqrt{k_{x}k_{z}}}{M}\ln\left(c_{\gamma=2/3}^{(1)}\frac{R_{x}}{\xi}\right),\label{eq:OmPrecFermi-1}
\end{equation}
 for the respective bosonic and fermionic cases. The other mode, which
we term \emph{the modulation mode}, is specific to the elements incorporated
via variations in the density. Its frequency $\Omega_{m}$, which
is higher than the precession frequency $\Omega_{p}$, has been obtained
through an analytical adiabatic approximation. Specifically, we have
found for $\Omega_{m}$ the following (bosonic and fermionic) expressions

\begin{equation}
\Omega_{m,B}=\frac{8}{3}\frac{\mu}{\hbar}\left(\left[2\ln\left(4\frac{R_{x}}{\xi}\right)-3\right]\left[2\ln\left(4\frac{R_{x}}{\xi}\right)-1-4\sqrt{\frac{k_{z}}{k_{x}}}\right]\right)^{-1/2}
\end{equation}
 and 

\begin{equation}
\Omega_{m,F}=2\frac{\mu}{\hbar}\left(\left[2\ln\left(2\frac{R_{x}}{\xi}\right)-2\right]\left[2\ln\left(2\frac{R_{x}}{\xi}\right)-4\sqrt{\frac{k_{z}}{k_{x}}}\right]\right)^{-1/2},
\end{equation}
 whose validity has been checked through numerical calculations. For
generic initial conditions, the global dynamics can be viewed as corresponding
to a combination of the two (component) modes of the system. The results
of the former description are recovered provided that the initial
conditions fulfill Ecs. (\ref{eq:VlocationEq1}) and (\ref{eq:VlocationEq2}):
only the precession mode is generated then.

iv) The present extension of the approach can be actually regarded
as a proof of consistency. The pertinence of improving the primary
ansatz by modifying the form of the density according to the precise
constraints imposed by the hydrodynamic formalism is clear. Since
the former order of approximation has been shown to account for salient
features of the dynamics, its robustness against a consistent modification
of the ansatz can be expected. The second-order character of the obtained
corrections confirms that argument. Following the same line of reasoning,
the physical character of the found second-order effects can be presumed.
However, their detection implies dealing with technical difficulties:
the observation of the additional (larger) frequency $\Omega_{m}$
requires higher experimental resolution. One cannot disregard that
the (uncontrolled) conditions for the emergence of the SVs structures
might correspond to the inhibition of the second mode. Moreover, we
should take into account that the potential resonance of that mode
with high-frequency collective excitations of the condensate could
activate a damping mechanism which can preclude its observation. The
analysis of the robustness of the second mode against dissipation
effects is left for future work.

\begin{figure}[H]
\centerline{\includegraphics{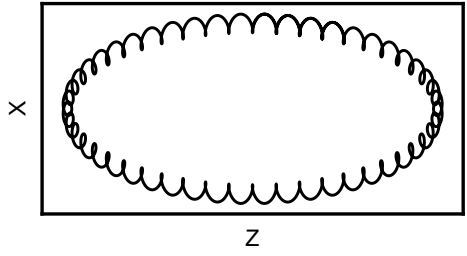}}\caption{An illustration of a SV trajectory as predicted by the first extension
of our basic approach. A perturbative high-frequency modulation of
the formerly obtained elliptical trayectory is observed. }
\end{figure}

\subsection{The effect of the condensate motion on the SV dynamics }

Now, we turn to analyze how the characterization of the vortex precession
is modified when degrees of freedom of the background fluid are taken
into account. The general procedure is illustrated by incorporating
the dipole and quadrupole modes into the description. Appropriate
to our objectives is the use of a variational ansatz $\Psi=\tilde{\rho}^{1/2}e^{iS}$
with the phase being given by 

\begin{equation}
S(\mathbf{r},t;x_{0},z_{0};a_{j},b_{jk})=S_{v}(x,z;x_{0},z_{0})-\frac{1}{\hbar}\bar{\mu}t+\delta+\sum_{j,k=x,z}a_{j}x_{j}+b_{jk}x_{j}x_{k},\label{eq:PhaseAnsatz2}
\end{equation}
 where, in addition to the vortex location and the normalization term
$\delta(t)$, we have included the set of parameters $a_{j}(t)$ and
$b_{jk}(t)$, which will allow dealing with the dipole and quadrupole
modes of the fluid. (Only the modes contained in the plane perpendicular
to the SV are considered.) Apart from entering the phase, those parameters
are present in the form of the density $\tilde{\rho}$, which is derived,
as indicated in the previous subsection, through the Euler-like (hydrodynamic)
equation that connects phase and density. Accordingly, we obtain

\begin{eqnarray}
\tilde{\rho} & = & -\frac{1}{g}\left[\frac{\hbar^{2}}{2M}\left|\boldsymbol{\nabla}S\right|^{2}+V_{ex}+\hbar\frac{\partial S_{v}}{\partial t}-\bar{\mu}+\frac{\dot{\delta}}{\hbar}+\sum_{j,k=x,z}\left(\dot{a_{j}}x_{j}+\dot{b}_{jk}x_{j}x_{k}\right)\right]\nonumber \\
 & = & \rho-\frac{1}{g}\left[\frac{\hbar^{2}}{2M}\left|\boldsymbol{\nabla}S\right|^{2}+\hbar\frac{\partial S_{v}}{\partial t}+\frac{\dot{\delta}}{\hbar}+\sum_{j,k=x,z}\left(\dot{a_{j}}x_{j}+\dot{b}_{jk}x_{j}x_{k}\right)\right],\label{eq:DensCmodes}
\end{eqnarray}
 where it is apparent that the terms $\dot{a_{j}}x_{j}$ account for
displacements of the center of the condensate, and the terms $\dot{b_{jk}}x_{j}x_{k}$
incorporate changes in the radii and reorientation of the axes. One
should notice the parallelism of this procedure with the method introduced
in Ref. \cite{Castin} to analyze the effect of modulations of the
trap frequencies on the fundamental state of a BEC. In that method,
it is the form of the density that is explicitly proposed since physically
supported conjectures can be made on it, the phase being subsequently
derived from the Euler-like hydrodynamic equation. In contrast, in
the present case, as it is the vorticity the characteristic of the
structure that is actually known, it is convenient to start the proposal
by modeling the phase. 

We proceed as before using Eq. (\ref{eq:BIGlagragian}) to build the
Lagragian function from the proposed ansatz, and, later on, to obtain
the Euler-Lagrange equations for the set of variational parameters.
In order to have a first global picture of the dynamics, we have worked
with a linearized version of the set of coupled equations. Within
this regime, all the integrals have been obtained analytically beyond
logarithmic accuracy to include zero order terms. The main implications
of the coupling of vortex and fluid coordinates are summarized in
the following points, where, to simplify the discussion we will not
refer to the modulation mode identified in the previous subsection.

i) The vortex precession can be tracked down in one of the emerging
normal modes. For the considered experimental conditions, given that
the inertia of the condensate is much larger than that of the SV structure,
the mixing of the former precession mode with the intrinsic condensate
modes, incorporated via the parameters $a_{j}(t)$ and $b_{jk}(t)$,
is negligible.  Indeed, the \emph{new version} of the precession mode
basically corresponds to the motion of the vortex relative to the
condensate. In contrast, there is a non-negligible displacement of
the mode frequency with respect to the formerly obtained $\Omega_{p}$.
That shift is specifically rooted in the coupling with the parameters
$a_{j}(t)$. Since its magnitude corresponds to a contribution of
zero order in the quotient $R_{x}/\xi$, it can be incorporated into
the functional form of $\Omega_{p}$ by modifying the effective zero-order
parameter, which will be denoted now $c_{\gamma}^{(3)}$. Accordingly,
the expressions of the precession frequencies corresponding respectively
to the bosonic and fermionic cases are written as 
\begin{equation}
\Omega_{p,B}=\frac{3}{4}\frac{\hbar}{\bar{\mu}}\frac{\sqrt{k_{x}k_{z}}}{M}\ln\left(c_{\gamma=1}^{(3)}\frac{R_{x}}{\xi}\right),\label{eq:OmegaPrec-1-2}
\end{equation}

\begin{equation}
\Omega_{p,F}=\frac{\hbar}{\bar{\mu}}\frac{\sqrt{k_{x}k_{z}}}{M}\ln\left(c_{\gamma=2/3}^{(3)}\frac{R_{x}}{\xi}\right).\label{eq:OmPrecFermi-1-1}
\end{equation}

ii) Among the resulting normal modes, one can also identify the dipole
and quadrupole modes of the condensate. The effect of the vortex on
those modes has been studied in previous work \cite{FetterVortexModes,StringariVortexModes}.
In agreement with the results of those studies, we observe that the
presence of the vortex affects the frequencies of the quadrupole modes
but leaves practically unchanged the frequencies of the dipole modes.
The implications of a stronger mixing of the vortex dynamics and the
condensate motion were analyzed in \cite{Fetter3}, where the setup
characteristics correspond to a reduction in the magnitude of the
inertia of the condensate relative to that of the vortex.

The results of the two studied extensions of the approach configure
a picture where, the corrections to the previously presented (primary)
description have a perturbative character.

\section{Application to the experiments}

In order to illustrate the applicability of the study to emulate the
results of the two considered experiments, we have incorporated into
our approach the two sets of force constants used in the practical
setups. Additionally, as amplitudes comparable to the TF radii were
reached in both experiments, we have considered a broad range of amplitudes
for the vortex motion. 

Figs. 2 and 3 depict our findings for the (fermionic) system studied
in \cite{Zwierlein1}. In both figures, the precession period of the
SV is displayed as a function of the chemical potential $\bar{\mu}$.
\textcolor{black}{(The precession period is expressed in units of
the period $T_{z}=2\pi/\omega_{z}$.)}\textcolor{green}{{} }Whereas
the results of Fig. 2 correspond to a linear regime, i.e., they are
applicable when the amplitude is sufficiently small for the approximation
$1-A_{xz}^{2}=1-X_{0}^{2}-Z_{0}^{2}\simeq1$ to be valid, Fig. 3 incorporates
nonlinear effects associated to increasing values of the amplitude.
Moreover, in order to illustrate how the predictions on the SV behavior
change as the sequence of extensions of the basic approach is applied,
partial results, corresponding to the different stages in our model,
are presented in Fig. 2. Namely, the dotted line reflects our primary
picture, where the period is given by Eq. (\ref{eq:OmPrecFermi}),
with the effective zero-order parameter taking the value $c_{\gamma}^{(0)}=1$.
This description already improves the results of early analyses through
the inclusion of anisotropy effects. Actually, at this stage, the
main characteristics of the experimental curves are approximately
reproduced. Still, observable corrections are obtained through the
inclusion of additional effects. The dashed line, which corresponds
to the first extension of the basic model, i.e., to the use of an
ansatz where the precise connection between phase and density is incorporated,
reflects a non-negligible modification of the period. A larger additional
shift (continuous line) is observed when the second extension of the
model is applied, i.e., when, in addition to the previous system components,
the coupling with the collective modes of the condensate is taken
into account. The values $c_{\gamma=1}^{(3)}=1.262$ and $c_{\gamma=2/3}^{(3)}=0.876$,
derived in our theoretical framework, were used to operatively incorporate
the contribution of zero-order terms at this level. In the two considered
regimes of the fermionic fluid, i.e., in the BEC side and in the unitary
regime, the agreement with the experimental results improves as the
description is generalized. Even so, it is the inclusion of nonlinearity
in our framework that constitutes the dominant correction to the primary
picture. As shown in Fig. 3, as larger amplitudes are reached, the
period significantly decreases approaching the experimental results
presented in \cite{Zwierlein1}. \textcolor{black}{(Note that the
open circles correspond to experimental results extracted from Fig.
3 of \cite{Zwierlein1}.)} It is also evident that although the agreement
is good in the whole range considered for the chemical potential,
the stronger dispersion of the experimental results in the unitarity
regime makes the comparison to be less conclusive in that region.
Actually, the need of a more detailed modeling of the system in that
range might be conjectured. Namely, a more precise approximation to
the size of the vortex core can be pertinent. In the same line, one
must take into account that the polytropic approximation to the equation
of state of the fluid, although appropriate to account for some significant
aspects of the dynamics, cannot be expected to give a complete description
of the system. 

\begin{figure}[H]
\centerline{\includegraphics{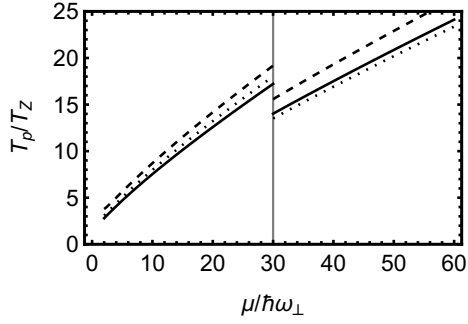}}\caption{The precession period of the SV in a fermionic fluid as a function
of the bulk chemical potential as given by the different approaches
developed in the study. (The amplitudes are small enough to guarantee
the applicability of a linear approximation.) The dotted line incorporates
the results obtained through the basic approach. The dashed line corresponds
to the first extension of the model. The continuous line represents
the results of the complete (linear) description.  The precession
period is expressed in units of the \textcolor{black}{period $T_{z}=2\pi/\omega_{z}$
corresponding} to the smallest of the trap frequencies, i.e., $\omega_{z}$.
Additionally, the chemical potential is written in units of $\hbar\omega_{\bot}\equiv\hbar\omega_{x}$.
\textcolor{black}{The figure illustrates the effect of the logarithmic
factor }$c_{\gamma}^{(i)}$\textcolor{black}{{} in each case. (The left
and right regions respectively correspond to the BEC and unitary regimes.) }}
\end{figure}

\begin{figure}[H]
\centerline{\includegraphics{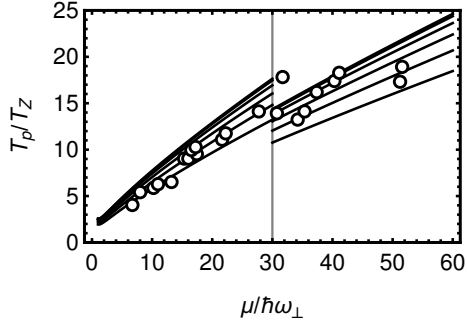}}\caption{The precession period of the SV in a fermionic fluid as a function
of the chemical potential for different amplitudes. The curves shift
down as the amplitude $A_{xz}$ takes larger values. (\textcolor{black}{$A_{xz}=0.,\,0.1,\,0.2,\,0.3,\,0.4,\,0.5$,
from top to bottom}.)\textcolor{black}{The open circles correspond
to experimental data extracted from Fig. 3 of \cite{Zwierlein1}.}
(Same units as in Fig. 2.)}
\end{figure}

Additional arguments on the importance of including nonlinearity in
the analyses can be extracted from Fig. 4, where the precession frequency
corresponding to the setup of \cite{Spielman1} is represented as
a function of the amplitude. Actually, the reproduction of the experimental
results (see the spectral analysis presented in Fig. 6 of \cite{Spielman1})
demands the introduction of nonlinear corrections into the model.
We recall that, in the early evaluation of the experiments of \cite{Spielman1},
the presence of SVs was merely conjectured since no confirmation through
direct observation techniques was feasible. The agreement of our predictions
with the experimental results supports the conjecture that the observed
structures are actually SVs.

\begin{figure}[H]
\centerline{\includegraphics{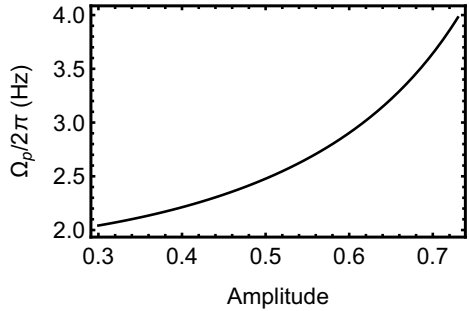}}\caption{The precession frequency corresponding to the setup of \cite{Spielman1}
as a function of the relative amplitude $A_{xz}$. }
\end{figure}

\section{Concluding remarks}

\textcolor{black}{Our study provides a theoretical framework for analyzing
the dynamics of SVs in trapped superfluids which extends former approaches
and allows clarifying recent experimental results. The unified approach
applied to SVs in bosonic and fermionic superfluids has served to
identify the common characteristics of the dynamics as simply rooted
in the superfluid character of both systems. It has been shown that,
in the regime where the hydrodynamic description is applicable, i.e.,
at scales much larger than the healing length, the only differential
aspect of the fluid statistics is a numerical factor, dependent on
the polytropic index, in the expression of the oscillation frequency.
The general correspondence of our results with those of the experiments
confirms the utility of the polytropic approximation to the equation
of state of the fermionic fluid as an operative method to uncover
basic aspects of the dynamics.}

\textcolor{black}{With respect to previous analyses of the considered
experiments, the study contains specific advances in understanding
the effect of the trap anisotropy, the relevance of nonlinearity to
the SV precession, and the implications of the coupling with collective
modes of the fluid. Indeed, the incorporation of a non-axisymmetric
trap in the applied model has served to trace the nontrivial dependence
of the oscillation frequency on the anisotropy transversal to the
vortical line. Moreover, nonlinearity has been found to be a central
component of the dynamics emergent in the implemented setups. We have
shown that the operatively defined inertial mass, apart from incorporating
characteristics of the structure and trapping, is amplitude dependent.
Additionally, the study has uncovered how the oscillation frequency
is shifted by the coupling with collective modes of the fluid. The
inclusion of those system components into our model implies a generalization
of former descriptions which has led us to obtain precise values for
the precession frequency, and, in turn, significantly improve  the
agreement with the experimental results.} Our whole approach enhances
the ground for the design of strategies of control. 

Apart from accounting for features observed in the experiments, our
analysis predicts the existence of fine details in the SV dynamics
associated to potentially realizable experimental conditions. Indeed,
the use of a variational ansatz where interrelated proposals for the
phase and density are consistently incorporated has revealed the presence
of a \emph{fine structure} in the previously identified SV trajectories.
The emergence of those fine details requires the implementation of
specific initial conditions. Although technically demanding, their
observation can be expected to be feasible given the significant advances
achieved in the control of the considered systems. \textcolor{black}{The
modified ansatz has also served to give a more complete characterization
of the vortex core. From our results, the validity of the basic modeling
of the core is confirmed.}

Some comments on potential extensions of the study are in order. No
border effects have been incorporated into our approach: straight
vortex lines have been considered. This simplification can be overcome
through an appropriate modification of the functional form of the
phase in the variational ansatz. The inclusion of dissipation effects,
which can have practical implications on the controlled preparation
of the systems and on the robustness of the predicted \emph{fine structure}
of the dynamics, is also pending. Moreover, we point out that the
present study, focused on the dynamics subsequent to the SVs formation,
does not complete the explanation of the experimental results. In
fact, the characterization of the whole decay sequence, starting from
PDS and ending with SVs is still required.  In this line, the inclusion
of additional structures in the decay process, like the intermediate
solitonic form predicted in \cite{MMateo1} and detected in \cite{Zwierlein2},
can be of great interest. Finally, it is worth stressing that alternative
theoretical approaches which go beyond the hydrodynamic description
of the fermionic system are needed to characterize the vortex dynamics
in the BCS regime.

\section*{Acknowledgments}

One of us (JMGL) acknowledges the support of the Spanish Ministerio
de Economía y Competitividad and the European Regional Development
Fund (Grant No. PID2019-105225GB-I00).

\end{document}